# CoNet: Borderless and decentralized server cooperation in edge computing

Ning Li, Member, IEEE, Xin Yuan, Zhaoxin Zhang, Jose Fernan Martinez

**Abstract-** In edge computing (EC), by offloading tasks to edge server or remote cloud, the system performance can be improved greatly. However, since the traffic distribution in EC is heterogeneous and dynamic, it is difficult for an individual edge server to provide satisfactory computation service anytime and anywhere. This issue motivated the researchers to study the cooperation between edge servers. The previous server cooperation algorithms have disadvantages since the cooperated region is limited within one-hop. However, the performance of EC can be improved further by releasing the restriction of cooperation region. Even some works have extended the cooperated region to multi-hops, they fail to support the task offloading which is one of the core issues of edge computing. Therefore, we propose a new decentralized and borderless server cooperation algorithm for edge computing which takes task offloading strategy into account, named CoNet. In CoNet, the cooperation region is not limited. Each server forms its own basic cooperation unit (BCU) and calculates its announced capability based on BCU. The server's capability, the processing delay, the task and calculation result forwarding delay are considered during the calculation. The task division strategy bases on the real capability of host-server and the announced capability of cooperation-servers. This cooperation process is recursive and will be terminated once the terminal condition is satisfied. The simulation results demonstrate the advantages of CoNet over previous works.

**Index Term-** Edge Computing, Server Cooperation, Task Offloading, Processing Delay.

## I. Introduction

The most important purpose of edge computing (EC) is to deploy cloud service close to mobile users as much as possible to speed up the task calculation at the greatest extent. In EC, by offloading the tasks to edge server or the remote cloud totally or partially, the performance of EC, such as the processing delay, the energy consumption, the network throughput, etc., can be improved greatly. However, since the traffic distribution in mobile edge networks is heterogeneous and dynamic, it is difficult for an individual edge server to provide satisfactory computation service anytime and anywhere. These issues motivated the researchers to study the cooperation among edge servers and remote cloud to extend the capability and improve the resource utilization of single server.

Two kinds of cooperation approaches have been proposed. The first one is the cooperation between edge server (ES) and remote cloud, such as the algorithms proposed in [1], [6], [7], [8], [11], [12], and [13]. In these algorithms, the single ES offloads tasks totally or partially to CS to extend its capability, which is effective. However, this approach has limited ability on dealing with heterogeneous and dynamic task distribution in EC, and the resource utilization is unbalance. For instance, when single server is overloaded, this approach is efficient; however, when the single ES is underloaded, the idle resources in this server cannot be used by other edge servers. This disadvantage can be addressed through cooperating between different ESs. In this kind of cooperation, when the load in one ES is heavy, this ES can offload the tasks to other ESs to extend its capability and improve the performance of EC, such as the algorithms proposed in [1], [2], [3], [4], [5], [9], [10], [14], [15], [16], etc. However, in these algorithms, the cooperation region is always limited within one-hop, i.e., each server can only cooperate with its one-hop neighbors. Thus, first, the restricted cooperation region cannot release the full potential of server cooperation. Second, considering the fact that the traffic distribution in EC is always regional, the adjacent ESs always have similar traffic distribution, i.e., idle or overloaded. Thus, the cooperating with one-hop ESs is shortsighted and the resource utilization in this cooperation manner is unbalance. For instance, if the edge servers in one area are overloaded, then cooperation with one-hop neighbor servers makes no contributions to improve the processing capability; on the contrary, if the edge servers in one area are idle, then the computation resource in these servers are wasted. Therefore, the restriction on cooperation region should be released to improve the performance of EC further.

To the best of our knowledge, the algorithm proposed in [17] is the first region-free server cooperation algorithm for EC. However, this algorithm has two disadvantages. First, it is more suitable for content caching rather than task offloading and calculating. In fact, it has already been applied in content caching instead of task offloading and calculating in [18] and [19]. Second, the task offloading strategy of mobile users, which is one of the core issues in EC, is not considered in this algorithm. Thus, by this approach, the mobile users still don't know when, where, and how many tasks should be offloaded to edge server. Moreover, due to the dynamics and diversity of offloading strategy and region-free server cooperation strategy, and they can affect each other greatly, these two strategies should be designed jointly rather than separately. However, this issue is not fully investigated by previous works. Additionally, considering the deep coupling between task offloading and the server cooperation, and the decentralized manner and dynamics in EC, achieving the joint optimization between these two items is challenging.

Therefore, in this paper, we propose a decentralized and borderless edge server cooperation algorithm for EC, named *cooperation network* (*CoNet*). Different from previous works, the main innovation of CoNet is that it achieves the joint optimization between task offloading and server cooperation successfully; moreover, for improving the processing capability and minimizing the processing delay as much as possible, the cooperation region is not limited. Specifically, in CoNet, each edge server constructs the basic cooperation unit (BCU) with its adjacent edge servers according to the construction principles that proposed in this paper. Then, each ES calculates its announced capability based on BCU, in which the server's capability, the processing delay, the task and calculation result forwarding delay are considered. The task offloading strategy and task division strategy between mobile users and ESs are decided by the announced capability. For guaranteeing the convergence and low-latency, we give the terminal condition for this cooperation. The cooperation process continues recursively and will be terminated once the terminal condition is satisfied. Moreover, we also prove that CoNet is $\varepsilon-$ approximation and $\varepsilon$ is small enough. The contributions of this paper are summarized as follows.

1. The main contribution of this paper is that we propose the decentralized and borderless ES cooperation approach for EC, i.e., CoNet. To the best of our knowledge, this is the first decentralized and borderless server cooperation approach which jointly optimizing the task offloading strategy and server cooperation strategy successfully.

2. In CoNet, first, we propose the basic cooperation unit (BCU) construction algorithm. In this algorithm, the cooperation principles between ESs are defined. Then, based on the constructed BCUs, we propose the announced capability calculation algorithm for ES, in which the server's capability, the processing delay, the task and calculation result forwarding delay are considered.

3. Since the task should be offloaded from mobile user to ES and divided between the cooperated servers, we propose the task offloading and division strategy based on the cooperation manner between ESs.

4. Finally, for guaranteeing the convergence and low-latency, we give the terminal condition for the cooperation. The cooperation process continues recursively and will be terminated once the terminal condition is satisfied. Additionally, we also prove that CoNet is $\varepsilon-$ approximation and $\varepsilon$ is small.

The rest of this paper is organized as follows. In Section II, we review the related works of server cooperation in edge computing. In Section III, we introduce the network model that used in this paper and state the problems that will be solved in this paper. Section IV introduces the details of CoNet. In Section V, we simulate and compare the performance of CoNet with previous works. Finally, in Section VI, we conclude our work in this paper.

## II. Related Works

For the multi-access edge computing, the main advantage is that the delay of task calculation in mobile user can be reduced significantly by offloading the calculation task from mobile users to the EC server. Thus, one of the fundamental issues in EC is the task offloading strategy. The task offloading is the first and hot research topic in EC, many excellent works have been proposed [20, 21]. For the task offloading in EC, only taking the task offloading into account is inadequate. Generally, the task offloading can be jointly considered with other performance metrics to improve the EC performance further, such as channel allocation [22], transmission power control [23, 24], etc. However, since the traffic distribution in mobile edge networks are heterogeneous and dynamic, it is difficult for an individual edge server to provide satisfactory computation service all the time. These challenges motivated the researchers to study the cooperation among edge servers which can be exploited to enhance EC performance via task offloading among them. The server cooperation in EC includes two aspects: 1) the cooperation between edge server and remote cloud, and 2) the cooperation between edge servers. The server cooperation is effective on improving the performance of content caching and task calculation in EC. In the following, we will introduce the related works on server cooperation in task calculation in detail.

Thus, some recent works have studied servers' cooperation for improving the effective of task calculation under various constraints. For instance, in [1], the authors investigate the edge-cloud and edge-edge cooperation to address the limitations of single ES and the nonuniform distribution of computation task arrivals among multiple ESs, in which two intelligent computation offloading algorithms based on Soft Actor Critic are proposed. In [2], the authors investigated the cooperation between multiple EC-BSs and propose a novel scheme to enhance the computation offloading service of an EC-BS through further offloading the extra tasks to other EC-BSs connected to it. In [3] and [4], by predicting the trajectory of vehicles, the authors propose to use the cooperation between servers to guarantee the service continuity in VANETs. In [5], the authors propose an online optimization strategy of EC server (ECS) computation task offloading with sleep control scheme to minimize the long-term energy consumption of the ECS network, in which the EC servers can cooperate with each other. In [6], the authors propose the dependency aware computation offloading for edge computing under the cooperation between edge-cloud cooperation. In [25], the authors propose a non-orthogonal multiple access assisted EC system, in which two near-far edge servers are cooperating to improve the performance of EC system. In [7], the authors propose that the edge-cloud cooperation can improve the performance of EC successfully. In [8], by using the edge-cloud cooperation, the authors propose a context-aware object detection algorithm for the vehicular network. The authors in [9] propose a workload management algorithm to allocate workloads among distributed data centers exploiting long term spatio-temporal diversities of water efficiency. In [10], the authors present a framework for dynamic service placement over time according to demand and resource price fluctuations based on game theoretic models. The authors in [26] develop a novel online small base station (SBS) peer offloading framework to optimize the long-term system delay under SBSs' long-term energy budget. To utilize the resources of the CS, the authors in [11] investigated the collaboration between cloud computing and edge computing to improve their computing efficiency by minimizing the weighted-sum latency of all MDs. The authors in [12] proposed an end-edge-cloud orchestrated computing architecture to improve the resource utilization and the QoS. The study in [13] investigated the issue of joint offloading decision and resource allocation for mobile cloud networks with a computing access point and a remote cloud center. However, the above studies did not consider the cooperation among multiple ESs. To address this issue, the studies in [14], [15] considered that multiple ESs in EC can offload computation tasks to each other. The authors in [16] proposed an Asynchronous-Advantage-Actor-Critic based real-time scheduler for stochastic edge-cloud environments allowing decentralized learning concurrently across multiple agents.

However, all the aforementioned server cooperation algorithms are only cooperated with its one-hop neighbors or the remote cloud server. They cannot release the full potential of the cooperation between ESs. Additionally, considering the fact that the traffic distribution in EC is always regional, the adjacent ESs always have similar traffic distribution, the cooperating with one-hop ESs is shortsighted and the resource utilization in this cooperation manner is unbalance.

## II. Network Model and Problem Statement

In EC, based on the performance requirements, the mobile users can offload the compute-intensive tasks to edge server that they belong to through wireless channel totally or partially. In this paper, we use $\beta$ to represent the task offloading ratio and $\beta \in [0\ 1]$. This means that the task in mobile users can be offloaded to EC server by arbitrary ratio. The edge servers can connect with neighbors with high-speed and high-bandwidth optical fibre. For each edge server, it can extend its capability and reduce the processing delay through cooperating with other edge servers. The cooperation between edge servers is dynamic and period. It changes with different network conditions and performance requirements. Different from previous works, the cooperation region in CoNet is not restricted. The delay calculation model, the offloading model, and the cooperation model are introduced as follows.

*A. The offloading model*

The offloading model includes: 1) the tasks offloading from mobile user to ES and from ES to its cooperated ESs; 2) the backhaul transmission of the calculation results between ESs.

*Task Offloading from mobile user to ES*. For mobile user $m$, when it has a computation-intensive task that needs to be calculated, it will decide: 1) offloading this task to the ES that it belongs to or calculating the task locally based on the processing delay in ES; 2) how many tasks can be offloaded to ES, i.e., the value of offloading ratio $\beta$. The task offloading between mobile user and ES is the same as that in previous works.

*Task Offloading between ESs*. However, in CoNet, the task offloading between different cooperated ESs is different from previous works. Before introducing the task offloading between cooperated ESs, we first introduce the cooperation model in CoNet. This model can be explained by Fig.1.

As shown in Fig.1, each server is expressed by a three-tuples $\langle s, Ca, \beta \rangle$, where $s$ means the ES and $Ca$ is the real capability of server $s$ itself. For instance, as shown in Fig.1, for the server in Level-1, its three-tuples is $\langle s_1^1, Ca_1^1, \beta_1^1 \rangle$. In Fig.1, $Tu$ means the tasks calculated by mobile user and $Ts$ is the task that processed by ES. Thus, we have:

$$T = Tu + Ts \quad (1)$$

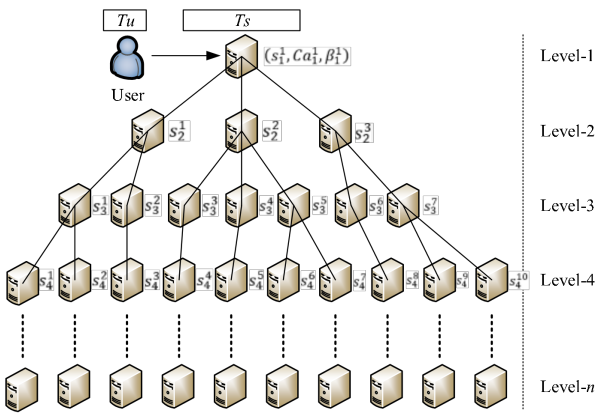

Fig.1 Cooperation Network

Next, we define the host-server and the cooperation-server in cooperation network as follows.

**Definition 1.** The host-server is the server who is in charge of dividing and forwarding tasks to other servers, i.e., the ESs who have child-nodes in Fig.1.

For instance, the $s_1^1$, $s_2^1$, $s_2^2$, and $s_2^3$ in Fig.1 are all host-servers.

**Definition 2.** The cooperation-server is the server who is the child-node of host-server, i.e., the ESs who have father-node in Fig.1.

For instance, the $s_2^1$, $s_2^2$, and $s_2^3$ in Fig.1 are also the cooperation-servers of $s_1^1$.

**Definition 3.** The host-server and its cooperation-servers form the basic cooperation unit (BCU).

For instance, the $s_1^1$, $s_2^1$, $s_2^2$, and $s_2^3$ form BCU with host-server is $s_1^1$ and the cooperation-servers are $s_2^1$, $s_2^2$, and $s_2^3$, denoted as $B^{s_1^1}(s_2^1, s_2^2, s_2^3)$.

Note that, in cooperation network, for each server, it can be host-server and cooperation-server simultaneously. For instance, for $s_2^1$, it is the host-server of $s_3^1$ and $s_3^2$, and it is also the cooperation-server of $s_1^1$. Moreover, in cooperation network, for the server in the first level, it can only be the host-server, such as $s_1^1$; for the servers in the final level, since they do not have child-nodes, they can only be cooperation-servers, such as the servers in Level-$n$ of Fig.1.

*B. The cooperation model*

Since the cooperation in CoNet is decentralized, each ES can only interact with its one-hop neighbors. However, this does not mean that the cooperation region in CoNet is one-hop. In the previous one-hop cooperation approach, the task $T_s$ can only be divided between the host-server and its one-hop cooperation-servers. In CoNet, different from the previous works, since the interaction between ESs in CoNet is recursive, this task can be arranged within multi-hops by decentralized manner. The details of the cooperation processes are presented as follows.

First, when the mobile user has a compute-intensive task to be processed, it will divide the task $T$ into $T = Tu + Ts$ and offload $Ts$ to ES ($s_1^1$ in Fig.2) that it belongs to according to the announced capability of this server. The announced capability is defined in Definition 4.

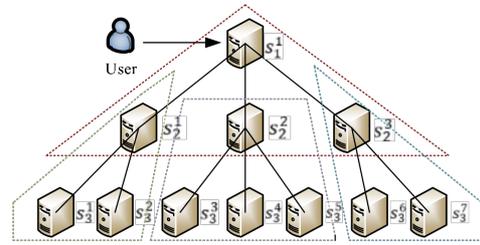

Fig.2. Announced Capability Calculation Model

**Definition 4.** The announced capability of server $s$ is the capability that server $s$ shows to other servers or mobile users except for its cooperation-servers, denoted as $c_s$. It includes both the real capability of server $s$ and the extended capability by its cooperation-servers.

Actually, the announced capability of server $s$ is the capability of the BCU in which the host-server is this server. For instance, as shown in Fig.2, for $B^{s_2^2}(s_3^3, s_3^4, s_3^5)$, the announced capabilities of the cooperation-servers are $c_{s_3^3}$, $c_{s_3^4}$, and $c_{s_3^5}$, respectively; the real capability of $s_2^2$ is $Ca_2^2$. Then, the announced capability of $s_2^2$ is:

$$c_{s_2^2} = Ca_2^2 + c_{s_3^3} + c_{s_3^4} + c_{s_3^5} \quad (2)$$

The calculation of $c_{s_3^3}$, $c_{s_3^4}$, and $c_{s_3^5}$ is the same as $c_{s_2^2}$, which will be introduced in detail in Section III. B.

When $s_1^1$ receives $Ts$, it will divide this task between itself and its cooperation-servers, i.e., $s_1^1$, $s_2^1$, $s_2^2$, and $s_2^3$, based on the real capability of itself $Ca_1^1$ and the announced capabilities of its cooperation-servers ($c_{s_2^1}$, $c_{s_2^2}$, and $c_{s_2^3}$).

Assuming that the task that divided to $s_2^2$ is $T_{s_2^2}$, then similar to the task dividing process in $s_1^1$, the $s_2^2$ will divide $T_{s_2^2}$ to its cooperation-servers based on the real capability of $s_2^2$ (i.e., $Ca_2^2$) and the announced capability of its cooperation-servers $s_3^3$, $s_3^4$, and $s_3^5$ (i.e., $c_{s_3^3}$, $c_{s_3^4}$, and $c_{s_3^5}$). Finally, the cooperation-servers of $s_1^1$ repeat this process recursively until the processing delay (including the calculation delay and the forwarding delay) satisfies the terminal condition. Based on this cooperation model, the cooperation region will be extended more than one hop. Note that the cooperation region in cooperation network is not fixed, it is optimized according to the task processing delay and task offloading strategy dynamically.

*C. The delay calculation model*

According to the cooperation model, the delay in CoNet includes three aspects. First, the calculation delay, which is the time used in ES for calculating task; second, the task forwarding delay, which is used for forwarding the divided tasks between the cooperated ESs; third, the calculation results forwarding delay, which is caused by forwarding the calculated result of task from ES back to mobile user.

*Capability.* The capability of mobile user is defined as the number of CPU cycles that used for processing 1-bit task, denoted as $Ca_m$. The $Ca_m$ can be calculated as:

$$Ca_m = \frac{1}{f_m} \quad (3)$$

where $f_m$ is the CPU frequency of mobile user $m$. For instance, if the task that processed in $m$ is $T$-bits, the time used on processing this task will be $TCa_m = \frac{T}{f_m}$.

The capability of ES is similar to that of mobile user, which is:

$$Ca_s = \frac{1}{f_s} \quad (4)$$

where $f_s$ is the CPU frequency of ES $s$, $Ca_s$ is the capability of $s$.

Note that considering the heterogeneity of devices, the capabilities of different mobile users and ESs are different, i.e., $f_m \neq f_s$, $f_{m_i} \neq f_{m_j}$ and $f_{s_i} \neq f_s$, $\forall i \neq j$.

*The processing delay.* In CoNet, the queue length in mobile users equals to 0. This means that the tasks generated by mobile users are processed immediately. For ES $s$, its queue length of unprocessed tasks is $q_s$. Thus, considering the capability of ES that defined in (4), the processing delay of $s$ that before task $k$ is processed, denoted as $D_s$, can be calculated as:

$$D_s = q_s Ca_s = \frac{q_s}{f_s} \quad (5)$$

Therefore, if the size of task $k$ is $l_k$, the time used for processing task $k$ is:

$$D_s^k = (q_s + l_k)Ca_s = \frac{q_s + l_k}{f_s} \quad (6)$$

*The task forwarding delay.* The task forwarding delay is the time used in ES for offloading tasks to its cooperation-servers. For ES $s$, there is a task forwarding queue, denoted as $q_s^{tf}$. For $s$, the task forwarding capability is defined as the number of CPU cycles that used for offloading 1-bit task to its cooperation-servers, denoted as $c_s^{tf}$. Thus, for $s$, the delay used to forward task to its cooperation-servers is:

$$D_s^{tf} = (q_s^{tf} + l_s^{tf})c_s^{tf} \quad (7)$$

where $l_s^{tf}$ is the size of task that offloaded from server $s$ to its cooperation-servers.

*The result forwarding delay.* Similarly, the result forwarding delay is defined as the time used for sending the calculated results from the cooperation-server $r$ back to its host-server. For $r$, the result forwarding queue length is denoted as $q_r^{rf}$. The result forwarding capability, which is defined as the number of CPU cycles that used for transmitting 1-bit calculation result back to its host-server, is denoted as $c_r^{rf}$. Moreover, for simplicity, we regard that the task forwarding capability equals to the result forwarding capability in the same ES, i.e., $c_r^{rf} = c_r^{tf}$. Thus, for $r$, the delay that used to forward the calculated results back to its host-server is:

$$D_r^{rf} = (q_r^{rf} + \gamma l_s^{tf})c_r^{rf} \quad (8)$$

where $\gamma$ means that for each task, if its size is $l$, the size of the calculated results of this task is $\gamma$ $(0 < \gamma < 1)$ time of $l$.

Note that, in this model, we omit the transmission delay in the communication link, because it is quite small compared with the processing delay and forwarding delay.

*D. Problem Statement*

Based on the network model that introduced above, three problems will be solved in CoNet. First, for each ES, how to form its BCU and optimize it dynamically. Second, how to divide the task between 1) mobile user and ES and 2) the cooperated servers, to minimize the processing delay. Third, when to terminate the cooperated task dividing and forwarding between cooperation-servers, i.e., the terminal condition of the cooperation process. Note that, in CoNet, considering the properties of EC, all the above issues should be addressed distributed. The problems can be formulated as follows.

Assuming that the task size is $T$, then the offloading ratio is $\beta = Ts/T$. In each cooperation region, all the ESs including the host-servers and cooperation-servers form the cooperation-server set, denoted as $S = \{s_1, s_2, s_3 ... s_n\}$, where $n$ is the number of ESs in this cooperation region. The set of the task offloading ratio in this cooperation region is $B = \{\beta_1, \beta_2, \beta_3 ... \beta_n\}$, where $\sum_{i=1}^{n} \beta_i = 1$.

Since the task $T$ is divided and processed in multiple devices parallelly (including mobile user and ESs), there are multiple parts of delays for task $T$. Thus, for minimizing the processing delay of task $T$, our purpose is to minimize the maximum delay of these cooperated servers and mobile user. Thus, this is a min-max combinatorial optimization problem. Based on the analysis above, the problem is solved in this paper can be expressed as:

$$\min \max \{D_m(\beta_m T), \max_i(D_{s_i}(\beta_{s_i}T) + T_{s_i}(\beta_{s_i}T))\} \quad (9)$$

$$\text{s.t.} \quad 1 \leq i \leq N \quad (9.a)$$
$$\beta_m + \sum_{i=1}^{N} \beta_{s_i} = 1 \quad (9.b)$$
$$0 \leq \beta_m \leq 1 \quad (9.c)$$
$$0 \leq \beta_{s_i} \leq 1 \quad (9.d)$$
$$0 < \gamma < 1 \quad (9.e)$$
$$q_{s_i} \leq q_{s_i}^{max} \quad (9.f)$$
$$q_{s_i}^{tf}, q_{r_i}^{rf} \leq q_f^{max} \quad (9.g)$$

where (9.a) means that the calculation should be limited in the cooperation region; the (9.b), (9.c), and (9.d) guarantee that the whole task $T$ should be processed among mobile user and cooperated servers; the (9.e) means the size of the computation result should smaller than that of the task; (9.f) and (9.g) guarantee that the queue length in ES should not exceed the maximum value. The above problem is a min-max problem. For the centralized approach, the solving is easy.

However, since our proposed algorithm is the decentralized manner, this problem is NP-hard (*Proved in Appendix A*).

IV. CoNet

In this section, we will introduce the details of CoNet. Moreover, we also analyze the theoretical performance of CoNet in this section in detail.

*A. The basic cooperation unit construction*

Based on the cooperation model that proposed in Section II.B, the first issue is to calculate the announced capability of ES. Because the task division between end devices and ESs is based on the announced capability of ES. As shown in Fig.2, the ESs exchange information with its one-hop cooperation-servers periodically. Note that, the one-hop cooperation-servers in CoNet are different from those in previous works. For instance, as shown in Fig.2, assuming that $s_3^2$, $s_3^3$, $s_3^4$, $s_3^5$, and $s_3^6$ are all one-hop neighbors of $s_2^2$, however, only $s_3^3$, $s_3^4$, and $s_3^5$ are the one-hop cooperation-servers of $s_2^2$. This is because, for each server, it may be the one-hop neighbor of more than one host-servers. For instance, $s_3^3$ is the one-hop neighbor of both host-server $s_2^1$ and $s_2^2$. So, how to construct the cooperation network should be investigated in detail. In the following, we will propose the BCU construction algorithm.

Firstly, during constructing the BCU, the servers who can connect with more than one host-servers should decide to select which one as its host-server.

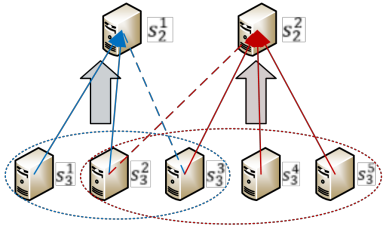

Fig.3. Host-server selection model

As shown in Fig.3, $s_2^1$ and $s_2^2$ are the two host-servers; $s_3^1$, $s_3^2$, and $s_3^3$ are the one-hop neighbors of $s_2^1$; $s_3^3$, $s_3^4$, and $s_3^5$ are the one-hop neighbors of $s_2^2$. Since $s_3^2$ and $s_3^3$ can connect to both $s_2^1$ and $s_2^2$, they need to decide to be the cooperation-servers of which host-server. For instance, we assume that $s_3^2$ chooses to be the cooperation-server of $s_2^2$, and $s_3^3$ chooses to be the cooperation-server of $s_2^2$. Thus, the one-hop neighbors and one-hop cooperation-servers are not the same items for host-servers. Specifically, in CoNet, the number of one-hop neighbors of host-server is no less than that of one-hop cooperation-servers, since not all the one-hop neighbors can be selected as the one-hop cooperation-servers of this host-server.

For selecting the host-server, three parameters of the host-server are considered: the processing queue length ($q_s$), the real capability ($Ca_s$), and the forwarding queue length ($q_s^{tf} + q_s^{rf}$). For minimizing the processing delay, the principle of host-server selection is that the selected host-server should have minimum $Ca_s$, $q_s$, and $q_s^{tf} + q_s^{rf}$. Note that the requirement that the selected host-server has minimum $q_s$ and $q_s^{tf} + q_s^{rf}$ is easy to be understood, because this can reduce the task calculation and result forwarding delay. The reason why we select the host-server which has minimum capacity is that: 1) this can improve the balance of network performance, and 2) can reduce the task processing delay.

However, in practice, finding a server in which the values of these three parameters are minimum simultaneously is always difficult or impossible. For solving this problem, we introduce the multi-attribution decision-making (MADM) approach into the host-server selection to choose the most appropriate host-server for the servers who connect to more than one host-servers.

Each host-server $s_i$ can be expressed by a parameter matrix $\left(Ca_{s_i}, q_{s_i}, q_f^{s_i}\right)$, where $Ca_{s_i}$ is the real capability, $q_i$ is the processing queue length, $q_f^{s_i} = q_{s_i}^{tf} + q_{s_i}^{rf}$ is the forwarding queue length. Based on the MADM algorithm and the performance matric, the utility of host-server can be calculated as:

$$U_i = \epsilon Ca_{s_i} + \delta q_{s_i} + \rho q_f^{s_i} \qquad (10)$$

where $\epsilon$, $\delta$, and $\rho$ are the weights of $Ca_{s_i}$, $q_{s_i}$, and $q_f^{s_i}$, respectively. For the MADM approach, the first important thing is to decide the weight for each parameter, i.e., the values of $\epsilon$, $\delta$, and $\rho$. The algorithms in [28] and [29] demonstrate that the weights of parameters can be calculated based on their variances, i.e., for the parameter whose variance is large, its weight should also be large; for the parameter whose variance is small, its weight is also small. Therefore, assuming that the variances of these three parameters are $v_{Ca}$, $v_q$, and $v_{qf}$, respectively, and considering the constraint that $\epsilon + \delta + \rho = 1$, we have:

$$\gamma = \frac{v_{Ca}}{v_{Ca}+v_q+v_{qf}} \qquad (11)$$
$$\delta = \frac{v_q}{v_{Ca}+v_q+v_{qf}} \qquad (12)$$
$$\rho = \frac{v_{qf}}{v_{Ca}+v_q+v_{qf}} \qquad (13)$$

Based on the utilities of the connected host-servers, for each EC server, if it has *n* neighbors, and the utilities of these *n* servers are $U_1$, $U_2$, …, $U_n$, then the utility of host server should satisfy:

$$U_i = \min_j\{U_i, j \in [1, n]\} \qquad (14)$$

The benefit of MADM based approach is simple, especially when the task flow is large, but the MADM approach takes both the processing delay and the forwarding delay into account. Thus, selecting the host server whose capability is the minimum can reduce the processing delay (*Proved in Appendix B*).

When the host server is selected, then all the servers will execute the above algorithm to form the cooperation network of the whole system. However, based on the above construction processed, it may form cooperation loop, as shown in Fig.4. The cooperation loop in the cooperation network is forbidden. Because the calculation of announcement capability of edge server will be incorrect if there is a cooperation loop. The announcement capability of EC server is defined as follows.

**Definition 5**. The announcement capability of EC server *s* is defined as the capability that *s* shows to other servers except for its cooperation servers; it includes both the capability of server *s* and the extended capability by its coordination servers.

For instance, as shown in Fig. 3, assuming that the cooperation servers of $s_2^2$ are $s_3^3$, $s_3^4$, and $s_3^5$, and the announcement capability of these three servers is $c_3^3$, $c_3^4$, and $c_3^5$, respectively. The real capability of $s_2^2$ is $C_2$. Then, the announcement capability of $s_2^2$ can be calculated as: $c_2^2 = C_2 + c_3^3 + c_3^4 + c_3^5$.

However, when forming the BCU, the cooperation between different BCUs may form a loop. The cooperation loop can affect the accuracy of announcement capability. As shown in Fig. 4, $s_2^1$ chooses $s_2^2$ as its host server, $s_2^2$ chooses $s_3^5$ as its host server, the host server of $s_3^5$ is $s_3^2$, the host server of $s_3^2$ is $s_3^4$, and $s_3^4$ chooses $s_2^1$ as its host server, then these cooperated servers form a cooperation loop, which is represented by the red arrows in Fig. 4. Once there is a loop, based on Definition 5, the announcement capability of $c_2^2$ will be $c_2^2 = C_2^2 + c_3^3 + c_2^1$. However, since $c_2^1 = C_2^1 + c_3^4$, $c_3^4 = C_3^4 + c_3^2$, $c_3^2 = C_3^2 + c_3^5$, and $c_3^5 = C_3^5 + c_3^1 + c_2^2$, $c_2^2$ also equals to $c_2^2 = C_2^2 + c_3^3 + C_2^1 + C_3^4 + C_3^2 + c_3^1 + c_2^2$, which is unacceptable. Therefore, once the host-server and cooperation-servers form a cooperation loop, the real capability and the announcement capability of one single server will be accounted repeatedly. The consequence is that the announcement capability that calculated under this scenario is inaccurate. Thus, the cooperation between host-server and cooperation-server should be loop-free. In the following, we propose the loop-free cooperation network construction algorithm.

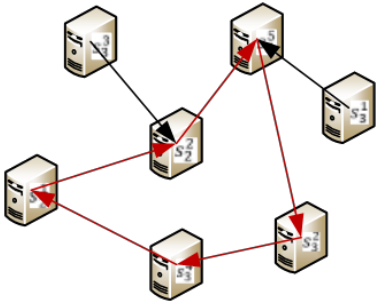

Fig. 4. An example of cooperation loop

In CoNet, each server matins a cooperation server matrix (CSM), denoted as $\{i|a,b,c\dots,k\}$, where $i$ means the host server, $\{a,b,c\dots,k\}$ are the cooperation-servers of $i$. For instance, as shown in Fig.4, the server $s_2^1$ and $s_3^3$ choose server $s_2^2$ as their host server, then the CSM of $s_2^2$ is $\{s_2^2|s_2^1, s_3^3\}$. The $s_2^2$ and $s_3^3$ are the cooperation-servers of $s_3^5$, so, the CSM of $s_3^5$ is $\{s_3^5|s_2^2, s_2^1, s_3^3, s_3^1\}$. Similarly, the CSM of $s_3^2$ is $\{s_3^2|s_3^5, s_2^2, s_2^1, s_3^3, s_3^1\}$, and the CSM of $s_3^4$ is $\{s_3^4|s_3^2, s_3^5, s_2^2, s_2^1, s_3^3, s_3^1\}$. Then, based on the host server of $s_3^4$, there are two situations: 1) if $s_3^4$ chooses server $s_2^1$ as its host server, the CSM of $s_2^1$ will be $\{s_2^1|s_3^4, s_3^2, s_3^5, s_2^2, s_2^1, s_3^3, s_3^1\}$, which forms a cooperation loop; 2) if $s_3^4$ selects the servers that are not shown in its CSM as the host server, there will be no loop.

Assuming that server $s_i$ has $n$ candidate host servers, which are $h_{s_i} = \{s_1, s_2, \dots, s_n\}$. The CSM of $s_i$ is $c_{s_i} = \{s_i|s_a, s_b, \dots, s_d\}$. Then, the Hamming distance between $h_{s_i}$ and $c_{s_i}$, i.e., $Hd_{s_i}$, can be calculated. For avoiding loop, the principle of loop-free are presented as follows.

1) If $Hd_{s_i} = 0$, $s_i$ chooses the host server from $h_{s_i}$ and the host server has the smallest utility that calculated by (J6).

2) If $Hd_{s_i} > 0$, first, finding the longest common subsequence between $h_{s_i}$ and $c_{s_i}$, noted as $l_{hc}^{s_i}$. Then, deleting the servers in $l_{hc}^{s_i}$ from $h_{s_i}$ and form the optimized candidate host server list $h'_{s_i}$. Finally, selecting the host server of $s_i$ from $h'_{s_i}$ and the host server has the minimum utility.

The detail of the greedy-based BCU construction algorithm is presented in Table 1.

Table 1. The greedy-based construction algorithm

| Algorithm 1: The greedy-based construction strategy |
|---|
| Input: Edge server set $S = \{s_1, s_2, s_3, \dots, s_n\}$; |
| Output: Constructed cooperation network of $S$; |
| 1. for $s_i \in S$, $i \in n$, finding its one-hop neighbors set $ne_i = \{s_a, s_k, \dots, s_j\}$; |
| 2. calculating the utilities of the servers in $ne_i$, noted as $U_i = \{u_a, u_k, \dots, u_j\}$; |
| 3. for server $s_i \in S$, $i \in n$, its CSM is $c_i = \{s_i|s_x, s_y, \dots, s_z\}$; |
| 4. the CSM of the servers in $ne_i$ is $c_k$, $k \in |ne_i|$; |
| 5. if $hd\{c_i, c_k\} = 0$ & $u_k$ is the smallest for $\forall k \in |ne_i|$; |
| 6. then $c_k = \{s_i|s_x, s_y, \dots, s_z, c_i\}$; |
| 7. otherwise |
| 8. if $hd\{c_i, c_k\} > 0$ |
| 9. then finding the longest common subsequence between $c_i$ and $c_k$, noted as $l_{ik}$; |
| 10. deleting $l_{ik}$ from $ne_i$ and repeating step 5 to step 9; |
| 11. until $hd\{c_i, c_k\} = 0$, $c_{k'} = \{s_i|s_x, s_y, \dots, s_z, c_i\}$ |

After that, for the cooperation between EC servers, three issues need to be investigated. First, how to calculate the announcement capability of each server based on the cooperation manner proposed above; second, how to divide the tasks between mobile user and the cooperated EC servers based on their announcement capability to minimum the processing delay; finally, what is the stop condition of this cooperation approach, since the cooperation must be stopped. In the following, we will investigate these issues in detail.

*B. Announced capability calculation*

Based on the BCU construction algorithm, the announced capability of host-server can be calculated. Note that, the announced capability calculation in CoNet is periodical, and it is nothing to do with the task offloading, forwarding, and processing. We take $B^{s_1^1}(s_2^1, s_2^2, s_2^3)$ in Fig.1 as an example to explain how to calculate the announced capability of host-server.

Since the announced capability of $s_1^1$ including 1) the capability of $s_1^1$ and its cooperation-servers, 2) the transmission delay between $s_1^1$ and its cooperation-servers, and 3) the transmission delay between $s_1^1$ and mobile user, for easy calculation, we redefined the capability of ES as follows.

**Definition 6**. The capability of ES $s$ when taking the delay (both the processing delay and the forwarding delay) into account is defined as the size of tasks that can be processed in unit delay $D_u$, denoted as $Cd_s$.

Note that this capability is different from that introduced in Section II.C, in which the capability equals to the reciprocal of CPU frequency. For instance, as $B^{s_1^1}(s_2^1, s_2^2, s_2^3)$ shown in Fig.1, $s_1^1$ is the host-server, $s_2^1$, $s_2^2$, and $s_2^3$ are cooperation-servers of $s_1^1$; then the delay in $s_1^1$ can be calculated as:

$$D_{s_1^1} = \frac{T_{s_1^1} + q_{s_1^1}}{f_{s_1^1}} + D_{s_1^1}^{tf} + D_{s_1^1-m}^{rf} \qquad (15)$$

where $T_{s_1^1}$ is the task that arranged to $s_1^1$, $f_{s_1^1}$ is the CPU frequency of $s_1^1$, $D_{s_1^1-m}^{rf}$ is the delay for transmitting the calculation result in $s_1^1$ from $s_1^1$ back to mobile user, $D_{s_1^1}^{tf}$ is the delay in $s_1^1$ for dividing and transmitting tasks to the servers in $B^{s_1^1}(s_2^1, s_2^2, s_2^3)$, $T_{B^{s_1^1}}$ is the task that offloaded

from mobile user to $s_1^1$. Similarly, the delay in $s_2^1$, $s_2^2$, and $s_2^3$ are:

$$D_{s_2^1} = \frac{T_{s_2^1}}{c_{s_2^1}^i} + D_{s_2^1-m}^{rf} \quad (16.a)$$

$$D_{s_2^2} = \frac{T_{s_2^2}}{c_{s_2^2}^i} + D_{s_2^2-m}^{rf} \quad (16.b)$$

$$D_{s_2^3} = \frac{T_{s_2^3}}{c_{s_2^3}^i} + D_{s_2^3-m}^{rf} \quad (16.c)$$

where $D_{s_2^2-m}^{rf}$ is the delay for transmitting the calculation result of $s_2^2$ from $s_1^1$ to mobile user, $c_{s_2^1}^i$ is the *ith* announced capability of $s_2^1$ (because the announced capability calculation is periodic). Note that the delay for the transmitting calculation result of $s_2^2$ from $s_2^2$ to $s_1^1$ is included in the announced capability of $B^{s_2^2}$. For $c_{s_1^1}^i$, $i \geq 0$; when $i=0$, $c_{s_1^1}^0 = f_{s_1^1}$. The calculation of $c_{s_1^1}^i$ is a recursive procedure when $i > 0$.

Then based on (7), (8), (15), and (16), the capability of ES can be calculated when $D_{s_1^1} = D_u$, $D_{s_2^1} = D_u$, $D_{s_2^2} = D_u$, and $D_{s_2^3} = D_u$, which are:

$$Cd_{s_1^1} = c_{s_1^1}^0 \left( D_u - D_{s_1^1}^{tf} - D_{s_1^1-m}^{rf} \right) - q_{s_1^1} \quad (17.a)$$

$$Cd_{s_2^1}^i = c_{s_2^1}^i \left( D_u - D_{s_2^1-m}^{rf} \right) \quad (17.b)$$

$$Cd_{s_2^2}^i = c_{s_2^2}^i \left( D_u - D_{s_2^2-m}^{rf} \right) \quad (17.c)$$

$$Cd_{s_2^3}^i = c_{s_2^3}^i \left( D_u - D_{s_2^3-m}^{rf} \right) \quad (17.d)$$

When the capabilities of these four servers are calculated, the announced capability of $s_1^1$ can be calculated as:

$$c_{s_1^1}^i = Cd_{s_1^1} + \sum_{j=1}^{3} Cd_{s_2^j}^i \quad (18)$$

where $c_{s_1^1}^i$ is the announced capability of $s_1^1$ and it is also the capability of $B^{s_1^1}(s_2^1, s_2^2, s_2^3)$.

Moreover, the values of the equations that shown in (17) can be expressed as:

$$Cd_{s_1^1} = \begin{cases} c_{s_1^1}^0 \left( D_u - D_{s_1^1}^{tf} - D_{s_1^1-m}^{rf} \right) + q_{s_1^1}, & \text{if } D_u > D_{s_1^1-m}^{rf} + D_{s_1^1}^{tf} + \frac{q_{s_1^1}}{c_{s_1^1}^0} \\ 0, & \text{if } D_u \leq D_{s_1^1-m}^{rf} + D_{s_1^1}^{tf} + \frac{q_{s_1^1}}{c_{s_1^1}^0} \end{cases} \quad (19.a)$$

$$Cd_{s_2^j}^i = \begin{cases} c_{s_2^j}^i \left( D_u - D_{s_2^j-m}^{rf} \right), & \text{if } D_u > D_{s_2^j-m}^{rf} \\ 0, & \text{if } D_u \leq D_{s_2^j-m}^{rf} \end{cases} \quad (19.b)$$

For $s_1^1$, if $D_u \leq D_{s_1^1-m}^{rf} + D_{s_1^1}^{tf} + \frac{q_{s_1^1}}{c_{s_1^1}^0}$, it means that due to the task forwarding delay, the result forwarding delay, and the processing queue delay are large enough, $s_1^1$ cannot process any data. Only when $D_u > D_{s_1^1-m}^{rf} + D_{s_1^1}^{tf} + \frac{q_{s_1^1}}{c_{s_1^1}^0}$, $s_1^1$ can process data. The meanings of (19.b) are the same as that of (19.a).

In the same BCU, the relationship between $D_u$ and the task forwarding delay, the result forwarding delay, and the processing queue delay is different. For instance, as $B^{s_1^1}(s_2^1, s_2^2, s_2^3)$ shown in Fig.2, there is one host-server ($s_1^1$) and three cooperation-servers ($s_2^1$, $s_2^2$, and $s_2^3$). For these ESs, the forwarding queue length and the task size that is divided into them are all different. Thus, in the same BCU, the constraints shown in (19) are different with different servers. Thus, for $B^{s_1^1}(s_2^1, s_2^2, s_2^3)$, we give

$$k_{s_2^j} = \frac{D_{s_2^j-m}^{rf}}{D_u} \quad (20.a)$$

$$k_{s_1^1} = \frac{D_{s_1^1}^{tf} + D_{s_1^1-m}^{rf} + \frac{q_{s_1^1}}{c_{s_1^1}^0}}{D_u} \quad (20.b)$$

where $k_{s_2^j}$ and $k_{s_1^1}$ are the parameters to evaluate the degree of forwarding delay in $B^{s_1^1}(s_2^1, s_2^2, s_2^3)$. When $0 < k_{s_1^1} < 1$ and $0 < k_{s_2^j} < 1$, it means that the servers in $B^{s_1^1}(s_2^1, s_2^2, s_2^3)$ can process data in unit delay. If $k_{s_2^j} \geq 1$ and $k_{s_1^1} \geq 1$, it means that the BCU cannot process any data in unit delay, i.e., the time that is needed to process one-bit data is larger than one unit delay. The larger $k_{s_i^j}$, the serious forwarding delay in $s_i^j$ is.

### C. Task division between cooperated servers

Based on the announced capability of ES and BCU, the task can be divided between these servers. Before introducing the task division algorithm, we first give the following conclusion.

The first constraint of task division between the cooperated servers is that it should make the delay in these servers and mobile user equal to each other. For instance, as shown in Fig.2, there are 11 servers and one mobile user. The task division should can guarantee that $D_{s_1^1} = D_{s_2^1} = D_{s_2^2} = D_{s_2^3} = D_{s_3^1} = D_{s_3^2} = D_{s_3^3} = D_{s_3^4} = D_{s_3^5} = D_{s_3^6} = D_{s_3^7} = D_m$, where $D_m$ is the delay of mobile user. This is easy to be understood since if the delay in one of the cooperation-server is lower than the others, then the tasks in other cooperation-servers can be re-arranged to this sever. Thus, the final convergent state of this cooperation approach is that the delay in all the cooperation-servers and mobile user is the same.

The second constraint is that the division should guarantee that all the tasks that offloaded from mobile user to servers should be arranged to the cooperated ESs. For instance, for $B^{s_1^1}(s_2^1, s_2^2, s_2^3)$, there should be $(1-\alpha)T = T_{s_1^1} + T_{s_2^1} + T_{s_2^2} + T_{s_2^3}$, where $\alpha$ is the ratio of task that calculated in mobile user, $T$ is the size of the whole task, $T_{s_1^1}$, $T_{s_2^1}$, $T_{s_2^2}$, and $T_{s_2^3}$ are the sizes of tasks that divided to $s_1^1$, $s_2^1$, $s_2^2$, and $s_2^3$, respectively. Based on these conclusions, we can give the task division algorithm as follows.

In Fig.2, assuming that the announced capability of $s_1^1$ is $c_{s_1^1}^i$, then the division of task $T$ between mobile user $m$ and $s_1^1$ can be calculated as:

$$\begin{cases} D_m = \frac{\alpha T}{f_m} + (1-\alpha)Tc_m^f \\ D_{s_1^1} = \frac{(1-\alpha)T}{c_{s_1^1}^i} \\ D_m = D_{s_1^1} \end{cases} \quad (21)$$

Based on (21), $\alpha$ can be calculated as:

$$\alpha = \frac{f_m\left(1 - c_m^f c_{s_1^1}^i\right)}{c_{s_1^1}^i + f_m\left(1 - c_m^f c_{s_1^1}^i\right)} \quad (22)$$

Based on (21) and (22), the delay is:

$$D_m = D_{s_1^1} = \frac{T}{c_{s_1^1}^i + f_m\left(1 - c_m^f c_{s_1^1}^i\right)} \quad (23)$$

In (23), there are two terms in the denominator: $c_{s_1^1}^i$ and $f_m\left(1 - c_m^f c_{s_1^1}^i\right)$. The $c_{s_1^1}^i$ is the announced capability of $s_1^1$;

$f_m \left(1 - c_m^f c_{s_1^1}^i\right)$ is the capability of mobile user except for the extra capability reduction caused by forwarding tasks to $s_1^1$. Thus, $f_m \left(1 - c_m^f c_{s_1^1}^i\right)$ can be regarded as the announced capability of mobile user. Thus, the meaning of (23) can be explained as the ratio of the whole size of task ($T$) to the whole announced capability of mobile user and cooperation-servers.

When $s_1^1$ receives $(1-\alpha)T$ tasks, it will divide these tasks between $s_1^1$, $s_2^1$, $s_2^2$, and $s_2^3$. This division can be calculated based on (9), (10), and (23). Moreover, in $B^{s_1^1}(s_2^1, s_2^2, s_2^3)$, $D_{s_1^1-m}^{rf}$, $D_{s_1^1}^{tf}$ and $D_{s_2^j-m}^{rf}$ can be calculated as follows:

$$D_{s_1^1-m}^{rf} = \left(q_{s_1^1}^{rf} + \gamma T_{s_1^1}\right) c_{s_1^1}^f \quad (24.a)$$

$$D_{s_1^1}^{tf} = \left(q_{s_1^1}^{tf} + (1-\alpha)T\right) c_{s_1^1}^f \quad (24.b)$$

$$D_{s_2^j-m}^{rf} = \left(q_{s_1^1}^{tf} + T_{s_2^j}\right) c_{s_1^1}^f \quad (24.c)$$

Then, based on (15), (17), (23), and (24), the tasks that divided to each server in $B^{s_1^1}(s_2^1, s_2^2, s_2^3)$ are:

$$T_{s_1^1} = \frac{c_{s_1^1}^0 \left[D_m - c_{s_1^1}^f(1-\alpha)T - \left(q_{s_1^1}^{rf} + q_{s_1^1}^{tf}\right)c_{s_1^1}^f\right] - q_{s_1^1}}{1 + \gamma c_{s_1^1}^f c_{s_1^1}^0} = \frac{c_{s_1^1}^0 g(T) - q_{s_1^1}}{1 + \gamma c_{s_1^1}^f c_{s_1^1}^0} \quad (25.a)$$

$$T_{s_2^1} = \frac{c_{s_2^1}^i \left[D_m - c_{s_1^1}^f q_{s_1^1}^{rf}\right]}{1 + \gamma c_{s_2^1}^i c_{s_1^1}^f} = \frac{c_{s_2^1}^i D_m - \theta_{s_1^1}}{1 + \gamma c_{s_2^1}^i c_{s_1^1}^f} \quad (25.b)$$

$$T_{s_2^2} = \frac{c_{s_2^2}^i \left[D_m - c_{s_1^1}^f q_{s_1^1}^{rf}\right]}{1 + \gamma c_{s_2^2}^i c_{s_1^1}^f} = \frac{c_{s_2^2}^i D_m - \theta_{s_1^1}}{1 + \gamma c_{s_2^2}^i c_{s_1^1}^f} \quad (25.c)$$

$$T_{s_2^3} = \frac{c_{s_2^3}^i \left[D_m - c_{s_1^1}^f q_{s_1^1}^{rf}\right]}{1 + \gamma c_{s_2^3}^i c_{s_1^1}^f} = \frac{c_{s_2^3}^i D_m - \theta_{s_1^1}}{1 + \gamma c_{s_2^3}^i c_{s_1^1}^f} \quad (25.d)$$

In (25.b), (25.c), and (25.d), $\theta_{s_1^1} = q_{s_1^1}^{rf} c_{s_1^1}^f$, $g(T) = D_m - c_{s_1^1}^f (1-\alpha)T - \sigma_{s_1^1}$, and $\sigma_{s_1^1} = \left(q_{s_1^1}^{rf} + q_{s_1^1}^{tf}\right) c_{s_1^1}^f$. However, for guarantee that $\beta_{s_1^1} + \beta_{s_2^1} + \beta_{s_2^2} + \beta_{s_2^3} = 1$, the ratios of the task that divided to the servers in $B^{s_1^1}(s_2^1, s_2^2, s_2^3)$ can be calculated as:

$$\beta_{s_1^1} = \frac{T_{s_1^1}}{T_{s_1^1} + T_{s_2^1} + T_{s_2^2} + T_{s_2^3}} \quad (26.a)$$

$$\beta_{s_2^1} = \frac{T_{s_2^1}}{T_{s_1^1} + T_{s_2^1} + T_{s_2^2} + T_{s_2^3}} \quad (26.b)$$

$$\beta_{s_2^2} = \frac{T_{s_2^2}}{T_{s_1^1} + T_{s_2^1} + T_{s_2^2} + T_{s_2^3}} \quad (26.c)$$

$$\beta_{s_2^3} = \frac{T_{s_2^3}}{T_{s_1^1} + T_{s_2^1} + T_{s_2^2} + T_{s_2^3}} \quad (26.d)$$

Moreover, for guaranteeing that $(1-\alpha)T = T_{s_1^1} + T_{s_2^1} + T_{s_2^2} + T_{s_2^3}$, the sizes of the task that divided to each server in $B^{s_1^1}(s_2^1, s_2^2, s_2^3)$ are $T'_{s_1^1} = \beta_{s_1^1}(1-\alpha)T$, $T'_{s_2^1} = \beta_{s_2^1}(1-\alpha)T$, $T'_{s_2^2} = \beta_{s_2^2}(1-\alpha)T$, and $T'_{s_2^3} = \beta_{s_2^3}(1-\alpha)T$. Thus, $T'_{s_1^1} + T'_{s_2^1} + T'_{s_2^2} + T'_{s_2^3} = (1-\alpha)T\left(\beta_{s_1^1} + \beta_{s_2^1} + \beta_{s_2^2} + \beta_{s_2^3}\right) = (1-\alpha)T$. But the recalculation of the task sizes that divided to the servers in $B^{s_1^1}(s_2^1, s_2^2, s_2^3)$ may make the delay in these servers does not equal to $D_m$. However, in the next section, we will prove that this variation is small.

From (16), we have the conclusion as follows.

**Corollary 1.** The mobile user can offload tasks to edge servers when the announced capability of ES (i.e., $c_{s_1^1}^i$) that the mobile user belongs to satisfies the constraints as: $c_{s_1^1}^i < \frac{1}{c_m^f}$ or $c_{s_1^1}^i > \frac{f_m}{f_m c_m^f - 1}$ and $f_m c_m^f > 1$.

*Proof.* See Appendix C. ∎

Based on Corollary 1, only when the announced capability of $s_1^1$ satisfies the constraints in Corollary 1, the mobile users can offload tasks to ESs. Otherwise, if $\frac{1}{c_m^f} < c_{s_1^1}^i < \frac{f_m}{f_m c_m^f - 1}$, all the tasks should be calculated locally, i.e., in mobile users.

For (25), we can find that with the increasing task size $T$, the variation of $T_{s_1^1}$, $T_{s_2^1}$, $T_{s_2^2}$, and $T_{s_2^3}$ are different. When $c_{s_1^1}^i < \frac{1}{c_m^f}$ or $c_{s_1^1}^i > \frac{f_m}{f_m c_m^f - 1}$ and $f_m c_m^f > 1$, i.e., the mobile user can offload tasks to edge servers, for $T_{s_1^1}$, if $\alpha > \frac{f_m\left(c_{s_1^1}^f - c_m^f\right)}{1 + f_m\left(c_{s_1^1}^f - c_m^f\right)}$, with the increasing $T$, the $T_{s_1^1}$ increases; if $\alpha < \frac{f_m\left(c_{s_1^1}^f - c_m^f\right)}{1 + f_m\left(c_{s_1^1}^f - c_m^f\right)}$, the $T_{s_1^1}$ reduces with the increasing $T$. Different from $T_{s_1^1}$, for $T_{s_2^1}$, $T_{s_2^2}$, and $T_{s_2^3}$, with the increasing $T$, all of them increase. This is easy to be understood. When $\alpha > \frac{f_m\left(c_{s_1^1}^f - c_m^f\right)}{1 + f_m\left(c_{s_1^1}^f - c_m^f\right)}$, it means that more tasks need to be calculated in mobile user and the value of $D_m$ is high. Thus, with the increasing $T$, the delay increases obviously in mobile users. Considering that the delay in mobile user and $B^{s_1^1}(s_2^1, s_2^2, s_2^3)$ are the same, thus, the size of the task in $s_1^1$ increases, too. When $\alpha < \frac{f_m\left(c_{s_1^1}^f - c_m^f\right)}{1 + f_m\left(c_{s_1^1}^f - c_m^f\right)}$, most of tasks are calculated in edge servers and the value of $D_m$ is small. With the increasing $T$, the delay increases slightly in mobile users because $\alpha$ is small. Therefore, the part of tasks that offloaded to the edge servers increases obviously. However, because $s_1^1$ is in charge of forwarding and dividing tasks to its cooperation-servers, if the task that offloaded to the ESs increases, the forwarding and dividing delay in $s_1^1$ increases, too. This will make the size of the task that can be processed in $s_1^1$ reduce since the value of $D_m$ is small. This conclusion is reasonable because when less part of task is offloaded to the edge servers, the delay will be decided mainly by mobile user and it is already high. So, it is not effective anymore for edge servers to cooperate with each other to reduce the processing delay. When a large part of task is offloaded to the edge server, the delay in mobile user will be very small. This means that the delay will be decided mainly by the edge servers. Thus, the cooperation between servers can reduce the processing delay greatly, which also means that $s_1^1$ should offload more tasks to its cooperation servers to reduce the delay in edge servers as much as possible.

### D. The terminal condition of cooperation

When the task division is achieved in $B^{s_1^1}(s_2^1, s_2^2, s_2^3)$, the servers in $B^{s_1^1}(s_2^1, s_2^2, s_2^3)$ except $s_1^1$ will act as the host-server and repeat the processes as $s_1^1$ recursively. However, this process cannot continue without termination. It should be terminated under certain conditions.

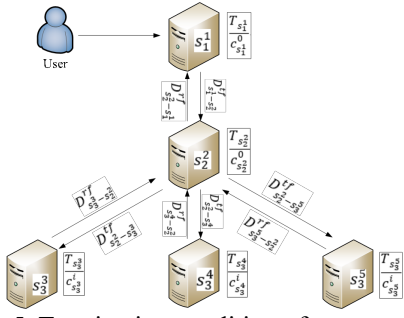

Fig.5. Termination condition of cooperation

In the cooperation network that shown in Fig.1, level-1 only contains one host-server $s_1^1$ that mobile user $m$ belongs to. The cooperation-servers of $s_1^1$ form the level-2, i.e., $s_2^1$, $s_2^2$, and $s_2^3$. When the servers in level-2 to be the host-servers, they will form their own BCUs. For instance, as shown in Fig.2, the $s_2^1$, $s_3^1$, and $s_3^2$ form the $B^{s_2^1}(s_3^1, s_3^2)$; the $s_2^2$, $s_3^3$, $s_3^4$, and $s_3^5$ form the $B^{s_2^2}(s_3^3, s_3^4, s_3^5)$. This process will continue until the termination condition is satisfied, which is introduced as follows.

For BCU $B^{s_i^j}$, the host-server is $s_i^j$ which is in level-$i$, its cooperation-servers are in level-($i$+1). Therefore, assuming that the task has been divided into all the cooperated servers, then based on the server cooperation model, (15), and (16), the delay of the servers in level-$i$ can be calculated as:

$$D_{s_i^j} = \frac{T_{s_i^j}}{c_{s_i^j}^0} + \sum_{n=i}^{2} D_{s_n^j - s_{n-1}^l}^{rf} + \sum_{n=i}^{2} D_{s_{n-1}^l - s_n^j}^{tf} + \mu_{s_i^j} \quad (27)$$

where $s_n^j$ is the $j$th servers in level-$n$, $s_{n-1}^l$ is the $l$th servers in level-($n$-1), and $s_{n-1}^l$ is the host-server of $s_n^j$, $D_{s_i^j - s_{i-1}^l}^{rf} = \left(q_{s_i^j}^{rf} + \gamma T_{s_i^j}\right) c_{s_i^j}$, $D_{s_i^j - s_{i-1}^l}^{tf} = \left(q_{s_i^j}^{tf} + T_{s_i^j}\right) c_{s_i^j}$, and $\mu_{s_i^j} = \frac{q_{s_i^j}}{c_{s_i^j}^0}$; the $\sum_{n=i}^{2} D_{s_n^j - s_{n-1}^l}^{rf}$ and $\sum_{n=i}^{2} D_{s_{n-1}^l - s_n^j}^{tf}$ are easy to be calculated by transmitting the value of task and result forwarding queue length from host-server to its cooperation-servers between two adjacent levels in cooperation network; $T_{s_i^j}$ is calculated by repeating the processes as $s_1^1$ recursively.

Based on (27), we can give the terminal condition as: *the process of cooperation will be terminated if and only if*:

$$D_m = \mu_{s_i^j} + \sum_{n=i}^{2} D_{s_n^j - s_{n-1}^l}^{rf} + \sum_{n=i}^{2} D_{s_{n-1}^l - s_n^j}^{tf} \quad (28)$$

This means that when the cooperation cannot improve the performance of processing delay anymore (i.e., when the accumulated forwarding delay is large enough), the cooperation should be terminated. Thus, in each level, the servers will judge whether this terminal condition is satisfied. If yes, the cooperation in this branch will be terminated; otherwise, the cooperation will continue.

Based on the terminal condition and (27), we can conclude that for different branches in cooperation network, the number of cooperation levels is also different. For instance, as shown in Fig.5, for the branches "$s_1^1 \to s_2^2 \to s_3^3$" and "$s_1^1 \to s_2^2 \to s_3^4$", the values of $D_{s_2^2 - s_1^1}^{rf}$, $D_{s_1^1 - s_2^2}^{tf}$, $D_{s_2^2 - s_1^1}^{rf}$, $D_{s_1^1 - s_2^2}^{tf}$, $D_{s_2^2 - s_1^1}^{rf}$, and $D_{s_1^1 - s_2^2}^{tf}$ relate to the size of task in $s_3^3$ and $s_3^4$ and the queue length in $s_1^1$ and $s_2^2$. However, the queue length of $s_1^1$ and $s_2^2$ is the same in these two branches, only the size of tasks in $s_3^3$ and $s_3^4$ are different. Assuming that $T_{s_3^3} > T_{s_3^4}$, then based on (27), $D_{s_3^3} > D_{s_3^4}$; thus, based on the terminal condition, when the cooperation in branch $s_1^1 \to s_2^2 \to s_3^3$ is stopped, the cooperation in branch $s_1^1 \to s_2^2 \to s_3^4$ may continue. Thus, in the cooperation network, the number of levels in each branch is different. Therefore, we give the approximation conclusion as follows.

Since the process of cooperation will be terminated if and only if the (28) is satisfied, let:

$$D_{min}^{rf} = min\{D_{s_i^j - s_{i-1}^l}^{rf}, D_{s_{i-1}^l - s_{i-2}^h}^{rf}, \ldots, D_{s_2^k - s_1^1}^{rf}\} \quad (29)$$

$$D_{max}^{rf} = max\{D_{s_i^j - s_{i-1}^l}^{rf}, D_{s_{i-1}^l - s_{i-2}^h}^{rf}, \ldots, D_{s_2^k - s_1^1}^{rf}\} \quad (30)$$

$$D_{min}^{tf} = min\{D_{s_{i-1}^l - s_i^j}^{tf}, D_{s_{i-2}^h - s_{i-1}^l}^{tf}, \ldots, D_{s_1^1 - s_2^k}^{tf}\} \quad (31)$$

$$D_{max}^{tf} = max\{D_{s_{i-1}^l - s_i^j}^{tf}, D_{s_{i-2}^h - s_{i-1}^l}^{tf}, \ldots, D_{s_1^1 - s_2^k}^{tf}\} \quad (32)$$

$$\mu_{max} = max\{\mu_{s_i^j}, \mu_{s_{i-1}^l}, \ldots, \mu_{s_1^1}\} \quad (33)$$

$$\mu_{min} = min\{\mu_{s_i^j}, \mu_{s_{i-1}^l}, \ldots, \mu_{s_1^1}\} \quad (34)$$

Then we have $\mu_{min} + (n-1)(D_{min}^{rf} + D_{min}^{tf}) < D_m < \mu_{max} + (n-1)(D_{max}^{rf} + D_{max}^{tf})$. Thus, the number of levels satisfies:

$$1 + \frac{D_m - \mu_{max}}{D_{max}^{rf} + D_{max}^{tf}} < n < 1 + \frac{D_m - \mu_{min}}{D_{min}^{rf} + D_{min}^{tf}} \quad (35)$$

*E. Theoretical analysis*

Before analyzing the performance of CoNet, we define the approximation ratio of this algorithm first.

**Definition 7**. The approximation ratio is defined as $\Delta = \frac{D_m}{D^*}$, where $D^*$ is the optimal task processing delay that calculated based on the centralized algorithm; moreover, if $\Delta \leq \varepsilon$, the proposed algorithm is $\varepsilon - $ approximation.

The $D^*$ can be calculated as follows. In the calculation of $D^*$, the number of cooperation-servers equals to that in the de-centralized manner; the queue length of task processing, task forwarding, and results forwarding in each server is known in advance. Then the value of $D^*$ can be calculated as the same process as (23), the only difference is that the $D^*$ is calculated centralized. However, since when the number of cooperation-servers is large, the calculation of $D^*$ is complex. Therefore, based on Definition 7 and (35), we give Corollary 2 as follows.

**Corollary 2**. The $\Delta$ is bounded; the lower and higher bounders of $\Delta$ are $\frac{D_m}{\mu_{max}+(n-1)(D_{max}^{rf}+D_{max}^{tf})}$ and $\frac{D_m}{\mu_{min}+(n-1)(D_{min}^{rf}+D_{min}^{tf})}$, respectively. Moreover, the CoNet is at most $\frac{D_{max}}{D_{min}} - approximation$.

*Proof.* See Appendix D. ∎

Since for the approximation ratio $\varepsilon$, the closer to 1, the better approximation is. Based on Corollary 2, one possible approach to improve the performance of CoNet is reducing the value of $\frac{D_{max}}{D_{min}}$ and makes it as closer to 1 as possible. Actually, the value of $\frac{D_{max}}{D_{min}}$ in CoNet is small. The main reason is that CoNet has an inherent load-balance mechanism. For instance, as shown in Fig.4, the values of $D_{s_3^5 - s_2^2}^{rf}$ and $D_{s_2^2 - s_3^5}^{tf}$ relate to the task and result forwarding queue length in $s_2^2$ and the size of task that divided to $s_3^5$. When the values of $D_{s_3^5 - s_2^2}^{rf}$ and $D_{s_2^2 - s_3^5}^{tf}$ are large, based on (16), the announced capability of $s_3^5$ and the size of task that divided to $s_3^5$ are small. This will reduce the values of $D_{s_3^5-s_2^2}^{rf}$ and $D_{s_2^2-s_3^5}^{tf}$. Based on this inherent load-balance mechanism, the value of $\frac{D_{max}}{D_{min}}$ cannot be very large. However, this analysis is only preliminary, due to the limitation of space, the deep analysis of load-balance and the approach for reducing $\frac{D_{max}}{D_{min}}$ will be left as our future work.

## V. Simulation and Discussion

In this section, we will evaluate and discuss the performance of the proposed CoNet in detail. The simulation environment, the baseline approaches, the variables and performance metrics are shown as follows.

*Simulation environment.* The simulation platform is an edge computing network with 200 ESs. Each ES connects to a random number of servers, this number is within (0 5]. The computational capacity of ES is uniformly distributed from 1 to 20 GHz. The task size is uniformly distributed in [1 8] Mbits. The computational capacity of ESs for transmitting tasks to ESs is randomly distributed from 5 to 15 GHz; the computational result of each task is 0.2 to the original size. The caching space in each ES is set to 10 Gbits. *Compared algorithms.* For demonstrating the advantages of CoNet, we compared the performance of CoNet with five previous works: EC-BS [2], CSACO and DSACO [1], all tasks that calculated locally, and the algorithm proposed in [27] (in which the cooperation only between mobile user and EC server, i.e., without EC server cooperation). *Variable sand performance metrics*. The variables used in this simulation are: the number of servers [60 200], the task size [1 8] Mbits, the task generation speed [0.1 1], and the CPU capability of edge server [1 20] GHz. The performance metrics used in this simulation are the network throughput and the processing delay.

### A. Performance under different numbers of servers

In this section, we evaluate the performance of network throughput, processing delay, and average cooperation distance under different number of servers. The variable is the number of EC servers, the task size is 4M, and the CPU capability is 8GHz. The results are presented in Fig.6, Fig.7, and Fig.8.

From Fig.6, we can find that the network throughput increases with the increasing of the number of EC servers except for the algorithm "Local". For the algorithm "Local", since all the tasks are calculated locally without offloading to EC server, its throughput is not affected by the number of EC servers. For the rest algorithms, the performance of CoNet is better than the other algorithms. The performance of CSACO and DSACO are similar and they all better than EC-BS. However, the DSACO is a little better than CSACO. The EC-BS is better than No-Cooperation. With the increasing of the number of EC servers, the increasing becomes slowly. For instance, for CoNet, when the number of EC servers increases from 60 to 120, the network throughput increases 32.6%; when the number of EC servers increases from 120 to 180, the network throughput increases 6.7%. The similar results can be found from the other algorithms. This is because when the number of servers is small, the workload is heavy, then once the number of servers increases, the workload reduces obviously. However, when the number of servers is larger enough, the reducing of workload is not as much as that when the number of servers is small. Thus, the increasing ratio reduces.

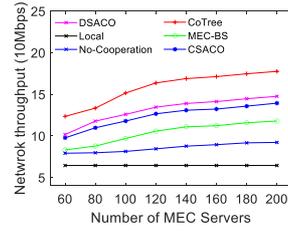
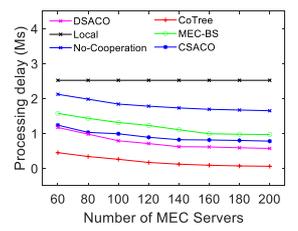

Fig.6 Network throughput under different number of servers

Fig.7 Processing delay under different number of servers

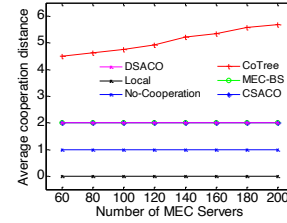

Fig.8 Average cooperation distance under different number of servers

The processing delay is shown in Fig.7. From Fig.7, we can conclude that the processing delay reduces with the increasing of the number of servers except for the algorithm "Local". The performance of CoNet is better than the other algorithms. With the increasing of the number of servers, the decreasing ratio reduces. For instance, when the number of servers increases from 60 to 120, the processing delay reduces from 0.45s to 0.17s; when the number of servers increases from 120 to 180, the delay reduces only 0.1s. The reason is similar to that in Fig.6.

The performance of average cooperation distance is presented in Fig.8. From Fig.8, we can find that except CoNet, the average cooperation distances of the other five algorithm keep constant with the increasing of the number of servers. This is because the cooperation distances in these five algorithms are fixed, while it is dynamic in CoNet. Note that since the algorithm "Local" does not offload task to EC server, we set its cooperation distance to 0. Similarly, the cooperation distance of "No-cooperation" is 1, and 2 in both DSACO, CSACO, and EC-BS. With the increasing of the number of servers, the cooperation distance in CoNet increases. For example, when the number of servers increases from 60 to 200, the average distance in CoNet increases 1.2. The reason is that with the increasing of the number of servers, the task size that assigned to each server reduces, then the processing and transmission delay between cooperation servers is reduced, which causes the increasing of cooperation distance.

### B. Performance under different task sizes

In this section, we evaluate the performance of these algorithms under different task sizes. The variable is the task size. The number of servers is 120, the CPU capability is 8GHz.

The Fig.9 shows the performance of network throughput under different task sizes. The network throughput of these algorithm decreases with the increasing of task size. Moreover, the performance of CoNet is better than the other algorithms. With the increasing of task size, the decreasing ratio increases. For instance, for the CoNet algorithm, when task size increases from 1 to 5, the network throughput increases 4.5%; when the number of EC servers increases from 5 to 8, the network throughput increases 12.7%. The similar results can be found from the other algorithms. The reason is that when the task size is small, once the task size increases, the system still has enough capability to process these tasks. However, when the task size is large enough, the processing and queueing delay increase obviously. Thus, the reducing ratio increases.

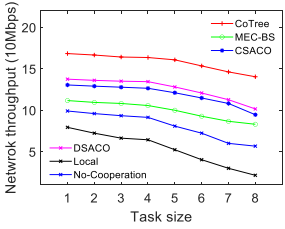
Fig.9 Network throughput under different task sizes

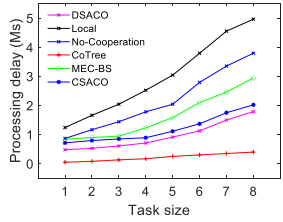
Fig.10 Processing delay under different task sizes

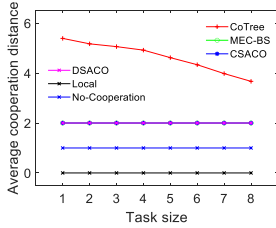
Fig.11 Average cooperation distance under different task sizes

The performance of processing delay is shown in Fig.10. In Fig.10, the processing delay increases with the increasing of task size. The CoNet is better than the other algorithms. With the increasing of the task size, the increasing ratio increases. For instance, in CoNet, when the task size increases from 1 to 4, the processing delay reduces from 0.05s to 0.17 s; when the task size increases from 4 to 8, the delay reduces only 0.23s. The reason is similar to that in Fig.9. Moreover, the increasing ratio in CoNet is the smallest. For example, when the task size increases from 1 to 8, the delay in CoNet increases 0.347s, while this value is 3.72s in "Local".

The performance of average cooperation distance is presented in Fig.11. In Fig.11, except for CoNet, the average cooperation distances in the other algorithms keep constant when the task size increases. The reason is the same as that introduced in Fig.8. Note that, similar to Fig.8, the average cooperation distances in DSACO, CSACO, and EC-BS are also coincident in Fig.11. With the increasing of task size, the cooperation distance in CoNet decreases. For instance, when the task size increases from 1 to 8, the average distance in CoNet reduces 1.72. This is because with the increasing of task size, the size of sub-tasks that assigned to each server increases, then the processing and transmission delay between cooperation servers increases, which causes the decreasing of average cooperation distance.

## C. Performance under different task generation speed

The performance of network throughput, processing delay, and average cooperation distance under different task generation speed is evaluated in this section. The variable is task generation speed. The task size is 4M, the CPU capability is 8GHz, and the number of servers is 120.

The Fig.12 shows the performance of network throughput under different task generation speed. From Fig.12, we can find that the network throughput reduces with the increasing of task generation speed. Moreover, the performance of CoNet is better than the other algorithms. With the increasing of the task generation speed, the network throughput reduction in CoNet is smaller than the other algorithms. For instance, when the task generation speed increases from 0.2 to 1, the throughput reduction in CoNet is 2.5; however, these values are 2.78, 2.66, 2.54, 4.95, and 2.97 in EC-BS, CSACO, DSACO, Local, and No-cooperation, respectively. This is because more servers cooperated in CoNet than the other algorithms. Thus, the CoNet has higher capability than the other algorithms.

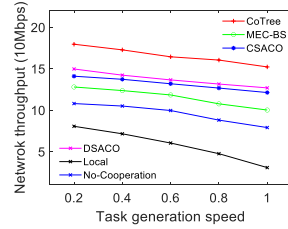
Fig.12 Network throughput under different task generation speed

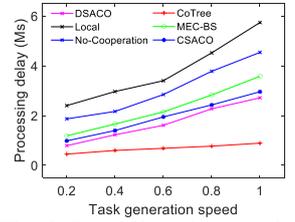
Fig.13 Processing delay under different task generation speed

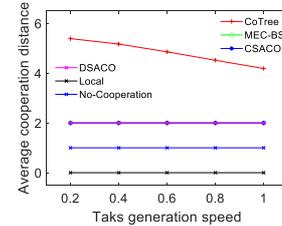
Fig.14 Average cooperation distance under different task generation speed

The performance of processing delay is shown in Fig.13. From Fig.13, we can conclude that the processing delay reduces with the increasing of the task generation speed. The performance of CoNet is better than the other algorithms. With the increasing of the task generation speed, the increasing ratio in the other algorithms is larger than that in CoNet. For instance, when the generation speed increases from 0.2 to 1, the processing delay increases 0.45. However, it increases 2.395, 1.97, 1.92, 3.34, and 2.67 in EC-BS, CSACO, DSACO, Local, and No-cooperation, respectively. The reason is similar to that in Fig.12.

The performance of average cooperation distance is presented in Fig.14. In Fig.14, except for CoNet, the average cooperation distances in the other five algorithms keep constant with the increasing of task size. The reason is the same as that introduced in Fig.8. Note that, similar to Fig.8, the results of DSACO, CSACO, and EC-BS are also coincident in Fig.14. With the increasing of task generation speed, the cooperation distance in CoNet decreases. For instance, when the task size increases from 1 to 8, the average distance in CoNet reduces 1.2. This is because with the increasing of task generation speed, the processing and transmission delay between cooperation servers increases. This will cause the reduction of cooperation distance.

## D. Performance under different CPU capabilities

In this section, we evaluate the performance of network throughput, processing delay, and average cooperation distance under different CPU frequencies. The variable is CPU capabilities. The task size is 4M, the number of servers is 120.

The Fig.15 shows the performance of network throughput under different CPU frequencies. In Fig.15, the network throughput increases with the increasing of CPU frequencies except for "Local". Moreover, the performance of CoNet is better than the other algorithms. With the increasing of CPU frequencies, the increasing ratio reduces. For instance, for CoNet, when the CPU frequencies increases from 2 to 8, the network throughput increases 53.8%; when the CPU frequencies increases from 12 to 16, the network throughput increases only 7.8%. The similar results can be found from the other algorithms. This is because when the CPU frequencies are small, the processing delay will be high, then if the CPU frequency increases, the delay reduces obviously. However, when the CPU frequency is large enough, the delay is already small enough; thus, the reduction is not obvious.

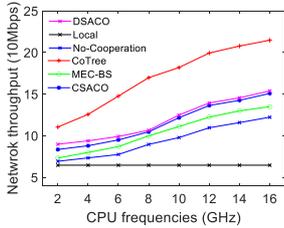

Fig.15 Network throughput under different CPU frequencies

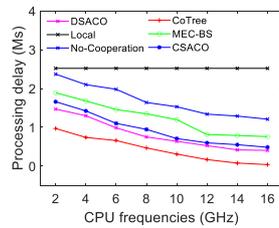

Fig.16 Processing delay under different CPU frequencies

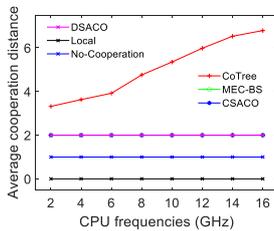

Fig.17 Average cooperation distance under different CPU frequencies

The performance of processing delay is shown in Fig.16. From Fig.16, we can conclude that the processing delay reduces with the increasing of CPU frequency except for "Local". The performance of CoNet is better than the other algorithms. With the increasing of CPU frequency, the decreasing ratio reduces. For instance, when the CPU frequency increases from 2 to 8, the processing delay reduces from 0.97s to 0.47 s; when the CPU frequency increases from 10 to 16, the delay reduces only 0.27s. The reason is similar to that in Fig.15.

The performance of average cooperation distance is presented in Fig.17. Except for CoNet, the average cooperation distances of the other five algorithm keep constant with the increasing of CPU frequencies. The reason is the same as that introduced in Fig.8. Note that, similar to Fig.8, the results of DSACO, CSACO, and EC-BS are also coincident in Fig.17. With the increasing of CPU frequencies, the cooperation distance of CoNet increases obviously. For instance, when the CPU frequency increases from 2 to 6, the average distance of CoNet increases 3.47. This is because with the increasing of CPU frequencies, the processing and transmission delay between cooperation servers are reduced greatly. This will cause the increasing of the cooperation distance.

## VI. Conclusions

In EC, the traffic distribution is heterogeneous and dynamic, it is difficult for an individual edge server to provide satisfactory computation service anytime and anywhere. Thus, in this paper, we propose a new decentralized and region-free ES cooperation approach for task offloading in EC, named cooperation network (CoNet). In CoNet, first, we propose the cooperation model. In this model, the cooperation region is not limited and each ES can cooperate with its one-hop cooperation-servers recursively until the terminal condition is met. Second, we propose the BCU construction principles in CoNet. In these principles, how the host-server cooperates with its cooperation-servers is defined. Then, based on these principles, we propose the announced capability calculation algorithm for edge server, in which the server's capability, the processing delay, the task and calculation result forwarding delay are considered. After that, we propose the task division algorithm between these cooperation-servers. Finally, based on the task division algorithm, we define the terminal condition of this cooperation model in CoNet. Moreover, we analyze the performance of CoNet by theoretical and simulation, which demonstrate that CoNet is better than previous works. However, due to the limitation of space, we only investigate the scenario which contains only one mobile user and multi ESs. The performance of CoNet under multi users will be presented in our future works.

## Appendix A

For the problem introduced above, we can give a particular case as follows. There are $N$ EC servers participate in the calculation of task $Ts$. The capabilities of these EC servers are known in advance, which is $Ca = \{Ca_1, Ca_2, ..., Ca_N\}$. The basic unit of task $Ts$ is $U_{Ts}$. This means that the $Ts$ can be divided into $m$ parts, where:

$$m = \frac{Ts}{U_{Ts}} \quad (A.1)$$

Moreover, we assume that $m \gg N$. This means that for each EC server, its received number of basic units of task $Ts$, denoted as $num$, may larger than 1, i.e., $0 \le num \le m$. Based on $num$, the ratio of task that divided to server $s_i$ can be calculated as: $\beta_i = num_{s_i}/m$. The purpose of this calculation model is to divide task $Ts$ to these N servers to calculate parallelly, and minimize the maximum processing delay of these N servers. For instance, assuming that two feasible solution are $num_1 = \{num_1^1, num_1^2, ... num_1^N\}$ and $num_2 = \{num_2^1, num_2^2, ... num_2^N\}$, then the processing delay for $num_1$ is $num_1^* = \max\{num_1^1, num_1^2, ... num_1^N\}$ and the processing delay for $num_2$ is $num_2^* = \max\{num_2^1, num_2^2, ... num_2^N\}$. Therefore, if $num_1^* < num_2^*$, then we regard that the solution $num_1$ is better than $num_2$. This is the same as the Shortest Parallel Scheduling Problem, which is a typical NP-complete problem.

Moreover, in the problem shown in (9), on one hand, the number of cooperated EC server is unknown, it should be decided during the task scheduling during these servers; on the other hand, since the strategy for each EC server should be decided distributed, there is not a centralized controller for making the task divided decision. Therefore, the problem presented in (9) is much more difficult to be solved than the particular case introduced above. Thus, if we can solve the problem presented in (9), then the particular case introduce above must can be solve, too. This means that if the optimal solution of (9) is $num^*$, then it must be the optimal solution of the particular case. Therefore, the problem presented in (9) is a NP-hard problem.

## Appendix B

The purpose of cooperation between EC servers is to extend the capability of single server and reduce the processing delay. On one hand, as shown in Fig. 4, assuming that the values of $q_s$ and $q_s^{tf} + q_s^{rf}$ in $s_2^1$ and $s_2^2$ are the same, and the capability of $s_2^1$ is smaller than that of $s_2^2$, then the cooperation server should choose to join $s_2^1$ to extend the capability of $s_2^1$ to minimize the processing delay of the tasks in $s_2^1$.

On the other hand, as shown in Fig. 4, assuming that the capability of $s_2^1$ and $s_2^2$ are $c_1$ and $c_2$, respectively, where $c_1 < c_2$; the tasks that calculated in $s_2^1$ and $s_2^2$ are $T_1$ and $T_2$, where $T_1 < T_2$. Based on the cooperation manner between EC servers, the relationship between $c_1$, $c_2$, $T_1$, and $T_2$ can be expressed as: $c_1/c_2 = T_1/T_2$, i.e., $T_2 = \frac{c_1}{c_2}T_1$. Moreover, we assume that the task and calculation result forwarding delay is $D$ in both these two servers. Then the delay in server $s_3^3$ can be expressed as: $d = \frac{T_3}{c_3} + D$. From this equation, we can find that the delay in $s_3^3$ relates to the size of the task that divided to $s_3^3$, i.e., $T_3$. Based on the cooperation manner, if $s_3^3$ is the cooperation server of $s_2^1$, then $T_3^1 = \frac{c_3}{c_1+c_3}T_1$; if $s_3^3$ is the cooperation server of $s_2^2$, then $T_3^2 = \frac{c_3}{c_2+c_3}T_2 = \frac{c_3}{c_2+c_3} \cdot \frac{c_1}{c_2}T_1$. Then $\frac{T_3^1}{T_3^2} = \frac{c_1(c_1+c_3)}{c_2(c_2+c_3)}$, since $c_1 < c_2$, we can conclude that $\frac{T_3^1}{T_3^2} < 1$, which equals to $T_3^1 < T_3^2$. This conclusion means that if $s_3^3$ is the cooperation server of $s_2^1$, its processing delay is smaller than that when $s_3^3$ is the cooperation server of $s_2^2$. ∎

## Appendix C

The $\alpha$ means the ratio of task that will be calculated in mobile user, thus, $0 \le \alpha \le 1$. As shown in (22), $\alpha = \frac{f_m\left(1-c_m^{rf}c_{s_1^1}^i\right)}{c_{s_1^1}^i + f_m\left(1-c_m^{f}c_{s_1^1}^i\right)}$. When $\alpha \ge 0$, there are two different situations.

First, $1 - c_m^{rf}c_{s_1^1}^i \ge 0$, which equals to $c_{s_1^1}^i < \frac{1}{c_m^f}$. In this situation, since $1 - c_m^{rf}c_{s_1^1}^i \ge 0$, $f_m\left(1 - c_m^{rf}c_{s_1^1}^i\right) \ge 0$ holds. Moreover, considering that $c_{s_1^1}^i > 0$, thus, $0 \le \alpha \le 1$ holds.

Second, $1 - c_m^{rf}c_{s_1^1}^i \le 0$, which equals to $c_{s_1^1}^i \ge \frac{1}{c_m^f} > 0$. In this situation, since $f_m\left(1 - c_m^{rf}c_{s_1^1}^i\right) \le 0$, if $\alpha \ge 0$ holds, there should be $c_{s_1^1}^i + f_m\left(1 - c_m^{f}c_{s_1^1}^i\right) \le 0$. Then, considering that $c_{s_1^1}^i \ge \frac{1}{c_m^f} > 0$, thus, we have $c_{s_1^1}^i > \frac{f_m}{f_m c_m^f - 1}$ and $f_m c_m^f > 1$. Moreover, since $\frac{1}{c_m^f} / \frac{f_m}{f_m c_m^f - 1} = \frac{f_m c_m^f - 1}{f_m c_m^f} < 1$, $0 \le \alpha \le 1$ holds.

## Appendix D

Since calculating the accurate $D^*$ is complex, we can estimate the boundary of $\Delta$ based on (35).

First, let 1) the task and result forwarding delay in all cooperation-servers equals to $D_{max}^{rf} + D_{max}^{tf}$, 2) the queueing delay of processing in all cooperation-servers equals to $\mu_{max}$; 3) the delay of all the servers in centralized approach equals to each other, thus, in this scenario, $D^*$ is $D^* = \mu_{max} + (n-1)(D_{max}^{rf} + D_{max}^{tf})$. This is the higher bound of delay in centralized approach.

Second, let 1) the task and result forwarding delay in all the cooperation-servers equals to $D_{min}^{rf} + D_{min}^{tf}$, 2) the queueing delay of processing in all cooperation-servers equals to $\mu_{min}$; 3) the delay of all the servers in centralized approach equals to each other, thus, $D^*$ is $D^* = \mu_{min} + (n-1)(D_{min}^{rf} + D_{min}^{tf})$, which is the lower bound of delay in centralized approach.

Thus, the boundary of $\Delta = \frac{D_m}{D^*}$ is:

$$\frac{D_m}{\mu_{max}+(n-1)(D_{max}^{rf}+D_{max}^{tf})} < \Delta < \frac{D_m}{\mu_{min}+(n-1)(D_{min}^{rf}+D_{min}^{tf})} \quad (D.1)$$

Moreover, since $\Delta \leq \varepsilon$, we can conclude that the CoNet is at most $\frac{D_m}{\mu_{min}+(n-1)(D_{min}^{rf}+D_{min}^{tf})} - approximation$. Based on (28), $D_{max}^{rf}$, $D_{max}^{tf}$, $D_{min}^{rf}$, and $D_{min}^{tf}$, there is $\mu_{min} + (n-1)(D_{min}^{rf} + D_{min}^{tf}) < D_m < \mu_{max} + (n-1)(D_{max}^{rf} + D_{max}^{tf})$. Therefore, considering the boundary of $\Delta$, we have:

$$\frac{\mu_{min}+(n-1)(D_{min}^{rf}+D_{min}^{tf})}{\mu_{max}+(n-1)(D_{max}^{rf}+D_{max}^{tf})} < \Delta < \frac{\mu_{max}+(n-1)(D_{max}^{rf}+D_{max}^{tf})}{\mu_{min}+(n-1)(D_{min}^{rf}+D_{min}^{tf})} \quad (D.2)$$

This can be expressed by:

$$\frac{(2n-1)D_{min}}{(2n-1)D_{max}} < \Delta < \frac{(2n-1)D_{max}}{(2n-1)D_{min}} \quad (D.3)$$

where $D_{min} = min\{D_{min}^{rf}, D_{min}^{tf}, \mu_{min}\}$ and $D_{max} = max\{D_{max}^{rf}, D_{max}^{tf}, \mu_{max}\}$, respectively. Thus, we have:

$$\frac{D_{min}}{D_{max}} < \Delta < \frac{D_{max}}{D_{min}} \quad (D.4)$$

Therefore, considering the fact that the approximation ratio $\varepsilon$ is larger than 1 and $\frac{D_{max}}{D_{min}} > 1$, then the boundary of $\varepsilon$ is:

$$1 \leq \varepsilon \leq \frac{D_{max}}{D_{min}} \quad (D.5)$$

where the $\varepsilon$ is at least 1 and at most $\frac{D_{max}}{D_{min}}$.